\documentclass[twocolumn,showpacs,preprintnumbers,amsmath,amssymb,superscriptaddress,nofootinbib]{revtex4}

\usepackage{subfigure}
\usepackage{amsmath,amssymb}
\usepackage{graphicx}
\usepackage{rotating}
\usepackage{bm}
\usepackage{axodraw}
\usepackage{color}

\definecolor{blue}{rgb}{0,0,1}

\definecolor{green}{rgb}{0,1,0}

\renewcommand{\hat}[1]{#1}

\newcommand{\ee}{\begin{equation}}
\newcommand{\eee}{\end{equation}}
\newcommand{\ea}{\begin{eqnarray}}
\newcommand{\eea}{\end{eqnarray}}

\definecolor{unterlegung}{rgb}{0.92,0.92,0.92}
\newlength{\importantlength}
\newcommand{\re}[1]{~(\ref{#1})}
\newcommand{\Cas}{C_2(\Nc)}

\renewcommand{\bar}{\overline}

\newcommand{\vv}{\ensuremath{v_{4}}}

\newcommand{\Nc}{\ensuremath{\textrm{N}_{\textrm{c}}}}

\newcommand{\fss}[1]{#1\!\!\!/}

\newcommand{\I}{\text{i}}

\newcommand{\casel}[2]{{\scriptstyle \frac{#1}{#2}}}
\newcommand{\Gk}{\Gamma_k}

\newcommand{\yb}{\bar{\psi}}
\newcommand{\Zy}{Z_{\psi}}

\newcommand{\pat}{\partial_t}

\newcommand{\SP}{\,(\text{S--P})}

\newcommand{\VAp}{\,(\text{V+A})}
\newcommand{\VAm}{\,(\text{V--A})}
\newcommand{\VAad}{\,(\text{V--A})^{\text{adj}}}
\newcommand{\VAn}{[2\!\VAad\!+({1}/{\Nc})\!\VAm]}

\newcommand{\gbar}{\bar{g}}
\newcommand{\Nf}{\ensuremath{\textrm{N}_{\text{f}}}}

\newcommand{\lp}{\hat{\lambda}_{+}}
\newcommand{\lm}{\hat{\lambda}_{-}}

\newcommand{\lsf}{\hat{\lambda}_{\sigma}}
\newcommand{\lva}{\hat{\lambda}_{\text{VA}}}

\newcommand{\blp}{\bar{\lambda}_{+}}
\newcommand{\blm}{\bar{\lambda}_{-}}

\newcommand{\blsf}{\bar{\lambda}_{\sigma}}
\newcommand{\blva}{\bar{\lambda}_{\text{VA}}}

\newcommand{\fsl}[1]{#1\!\!\!\!/}
\newcommand{\lF}{l_1^{\text{(F)},4}}
\newcommand{\lFB}{l^{\textrm{(FB)},4}_{1,2}}
\newcommand{\lFBo}{l^{\textrm{(FB)},4}_{1,1}}

\newcommand{\LUV}{\Lambda_{\text{UV}}}
\newcommand{\xsb}{$\chi$SB}

\definecolor{red}{rgb}{1,0,0}

\begin{document}

\author{Holger Gies}
\affiliation{Institut f\"ur theoretische Physik, Philosophenweg 16,
  69120 Heidelberg} 
\author{Joerg Jaeckel}
\affiliation{Deutsches Elektronen Synchrotron, Notkestrasse 85,
22607 Hamburg}
\preprint{HD-THEP-05-16}
\preprint{DESY-05-126}
\title{Chiral phase structure of QCD with many flavors}

\begin{abstract}
We investigate QCD with a large number of massless flavors with the
aid of renormalization group flow equations. We determine the critical
number of flavors separating the phases with and without chiral
symmetry breaking in SU($\Nc$) gauge theory with many fermion
flavors. Our analysis includes all possible fermionic interaction
channels in the pointlike four-fermion limit. Constraints from gauge
invariance are resolved explicitly and regulator-scheme dependencies
are studied. Our findings confirm the existence of an $\Nf$ window
where the system is asymptotically free in the ultraviolet, but
remains massless and chirally invariant on all scales, approaching a
conformal fixed point in the infrared. Our prediction for the critical
number of flavors of the zero-temperature chiral phase transition in
SU(3) is $\Nf^{\text{cr}}=10.0\pm
  0.29\text{(fermion)}\genfrac{}{}{0pt}{}{+1.55}{-0.63}\text{(gluon)}$, with the
errors arising from approximations in the fermionic and gluonic
sectors, respectively.
\end{abstract}
\maketitle

\section{Introduction}

In order to investigate the dynamics of strongly interacting gauge
systems, it has often been a successful strategy to deform the desired
system into a more accessible one. For instance, the approximation of
a continuum gauge theory by a finite lattice gauge theory can be
viewed as such a deformation. In the same spirit, the addition of
more symmetries as in supersymmetric versions of gauge
theories represents such a more tractable deformation. In the present
work, we consider the deformation of SU($\Nc$) gauge theories by many
massless fermion flavors. Since light fermions have the perturbative
tendency to screen long-range forces, a large fermion flavor number
$\Nf$ has the potential to move the system towards weaker gauge
interaction for which analytic tools are more powerful.

The relevance of fermionic screening at large $\Nf$ is ultimately
observable for $\Nf>\Nf^{\text{a.f.}}:=\frac{11}{2} \Nc$ where
asymptotic freedom is lost. But even for smaller $\Nf$, fermionic
screening is first signaled by the second $\beta$ function
coefficient of the gauge coupling, reversing its sign for
$\Nf>\frac{34\Nc^3}{13\Nc^2-3}$. This sign change induces
a second zero of the perturbative $\beta$ function, implying an
infrared (IR)
attractive fixed point of the gauge coupling $\alpha_\ast>0$. As was
first argued in \cite{Banks:1981nn}, the fixed-point value
$\alpha_\ast$ is small for $\Nf\lesssim\Nf^{\text{a.f.}}$, supporting
the possibility of a perturbative analysis. If this fixed point
persists on all scales, the system approaches a conformally invariant
limit in the deep IR, keeping massless quark and gluon excitations
in the spectrum. This scenario can collapse owing to nonperturbative
phenomena such as the spontaneous break-down of chiral symmetry. The
latter renders the fermions massive, implying their decoupling at low
scales. Fermionic screening properties are thus switched off in the
deep IR, and the system is characterized by strongly coupled glue and the
Goldstone bosons of chiral symmetry breaking (\xsb).  

Since \xsb\ is triggered by the strength of the gauge interactions
which in turn depends on the amount of fermionic screening, we expect
the conformal scenario to hold for large $\Nf$ sufficiently close to
$\Nf^{\text{a.f.}}$. The broken phase is supposed to occur for smaller
$\Nf$ with a critical $\Nf^{\text{cr}}$ separating the two phases. In
fact, evidence for this phase structure has been collected from
various methods, including (improved) ladder truncated Dyson-Schwinger
equations \cite{Miransky:1996pd,Appelquist:1996dq,Harada:2003dc},
lattice QCD \cite{Iwasaki:2003de}, anomaly-induced effective
potentials \cite{Sannino:1999qe}, and instanton models
\cite{Appelquist:1997dc}, with estimates for the critical number of
flavors ranging from $\Nf^{\text{cr}}\simeq 5$ \cite{Harada:2000kb}
to $\Nf^{\text{cr}}\simeq12$ for SU(3) gauge theory.

An analysis of the quark-scattering amplitude using the functional
methods \cite{Miransky:1996pd,Appelquist:1996dq} particularly reveals
that the nature of the phase transition, though continuous, is not
conventionally second order.\footnote{Owing to the different conformal
properties of the system on each side, the transition is often
referred to as a conformal phase transition \cite{Miransky:1996pd}.}
This is most prominently visible in the fact that there appear to be
no light scalar states in terms of which an effective theory could be
constructed on the symmetric side ($\Nf\gtrsim\Nf^{\text{cr}}$) of the
phase transition
\cite{Miransky:1996pd,Appelquist:1996dq,Chivukula:1996kg}. Any attempt
at constructing a Ginzburg-Landau-type of effective potential for the
chiral order parameter hence produces discontinuities in the potential
parameters. On the other hand, the order parameter in the form of the
chiral condensate, i.e., the minimum of this would-be potential,
changes continuously across the phase transition.

In this work, we analyze the phase structure of many-flavor QCD with
the aid of the functional renormalization group (RG). Beyond its
conceptual simplicity, the approach operates in continuous spacetime,
supports an explicit implementation of chiral symmetry, and includes a
resummation of beyond-ladder diagrams already in simple
truncations. Moreover, we monitor gauge invariance with the aid of
Ward-Takahashi identities and study regulator-scheme
dependencies. Since the critical number of flavors is a universal
quantity, the scheme-dependencies are a direct measure of the
approximation uncertainties of our nonperturbative truncation. Apart
from a quantitative prediction for $\Nf^{\text{cr}}$, our results
reveal further facets of the nature of the phase transition. Across
the phase transition, we particularly observe discontinuities in the
effective fermionic self-interactions which are generated by the RG
flow. 

For our analysis, we employ the functional RG formulated in terms of a
flow equation for the effective average action $\Gamma_k$
\cite{Wetterich:1993yh},
\begin{equation}
\pat\Gamma_k=\frac{1}{2}\, \text{STr}\, \pat R_k\,
(\Gamma_k^{(2)}+R_k)^{-1}, \quad t=\ln \frac{k}{\LUV}, \label{ERG}
\end{equation}
serving as an alternative definition of quantum field theory.
$\Gamma_k$ is a free-energy functional that interpolates between the
bare action $\Gamma_{k=\LUV}= S$ and the full quantum effective action
$\Gamma=\Gamma_{k=0}$. Here, $R_k$ denotes a regulator function that
specifies the details of the momentum-shell integrations, the
variation of which will provide us with an error estimate of our
approximations. We devote Sect. \ref{flow} to a discussion of our
approximation scheme, which represents a truncation of the full
theory to the relevant operators for the present problem.  Results and
conclusions are presented in Sects. \ref{results} and \ref{summary}.

\section{Flow equations}\label{flow}

Our primary aim is a reliable determination of the critical flavor
number $\Nf^{\text{cr}}$. For this, the approach of the phase
transition from the symmetric side is technically advantageous, since
the relevant degrees of freedom are expected to remain the same during
the flow from the ultraviolet (UV) to the infrared (IR): quarks and gluons. Therefore, we
solve the flow equation \eqref{ERG} in a truncated subspace of all
possible action functionals. 

In addition to standard gauge and fermion sectors, we include all possible
pointlike four-fermion interactions that are compatible with
$\textrm{SU}(\Nc)$ gauge symmetry and a chiral
$\textrm{SU}(\Nf)_{\textrm{L}}\times \textrm{SU}(\Nf)_{\textrm{R}}$
flavor symmetry for $\Nf$ fermion species\footnote{We note that only
  the four-fermion interactions are manifestly invariant under local
  gauge transformations for all possible choices of the couplings.
  Gauge invariance of the remaining terms is governed by (modified)
  Ward-Takahashi identities as discussed for the present system in
  \cite{Gies:2003dp}.},
\begin{eqnarray}
\nonumber \Gk\!\!\!&=&\!\!\!\!\int
\yb(\I\Zy\fss{\partial}+Z_{1}\bar{g}\fsl{A})\psi
+\frac{Z_A}{4}F^{\mu\nu}_z F_{\mu\nu}^z
    +\frac{(\partial_{\mu}A^{\mu})^2}{2\alpha}
\label{equ::truncsym}\\\nonumber &&\!\!+\frac{1}{2} \Big[
  Z_{-}\blm\!\VAm + Z_{+}\blp \!\VAp
  +Z_\sigma\blsf\! \SP
  \\
  &&\quad\quad+Z_{\textrm{VA}}\blva \VAn \Big].
\end{eqnarray}
Here $A_\mu=A^z T^z$, $F_{\mu\nu}=F_{\mu\nu}^z T^z$ denote the
nonabelian gauge potential and field strength (the inclusion of a
ghost sector is tacitly assumed). The gauge-field kinetic
term is accompanied by a $k$-dependent wave-function renormalization
$Z_A$, the fermionic one by $\Zy$. Similarly, $Z_{1}$,
 $Z_{+}$, $Z_{-}$, $Z_{\sigma}$ and
$Z_{\textrm{VA}}$ are the vertex renormalizations, whereas
$\gbar$, $\bar{\lambda}$ denote the bare couplings. Renormalized
dimensionless couplings are defined by
\begin{equation}
g=\frac{\bar{g}Z_{1}}{Z_A^{{1}/{2}}Z_{\psi}}, \quad
\hat{\lambda}_i=\frac{Z_{i}k^{2}\bar{\lambda_i}}{\Zy^2}.
\end{equation}
We work in the Landau gauge, $\alpha=0$, which is known
to be a fixed point of the renormalization group
\cite{Ellwanger:1995qf} and has the additional advantage that the
fermionic wave function is not renormalized in our truncation,
such that we can choose $\Zy=1$.

The four-fermion interactions can be classified according to their
color and flavor structure. Color and flavor singlets are
\begin{eqnarray}
\VAm&=&(\yb\gamma_\mu\psi)^2 + (\yb\gamma_\mu\gamma_5\psi)^2,
\\
\VAp&=&(\yb\gamma_\mu\psi)^2 - (\yb\gamma_\mu\gamma_5\psi)^2 ,
\end{eqnarray}
where color ($i,j,\dots$) and flavor ($a,b,\dots$) indices are
contracted pairwise, e.g., $(\yb\psi)\equiv (\yb_i^a \psi_i^a)$.
The remaining operators have non-trivial color or flavor structure
\begin{eqnarray}
\SP&=&(\yb^a\psi^b)^2-(\yb^a\gamma_5\psi^b)^2
\\\nonumber
&\equiv&
   (\yb_i^a\psi_i^b)^2-(\yb_i^a\gamma_5\psi_i^b)^2,\nonumber\\
\VAad&=&(\yb \gamma_\mu T^z\psi)^2 + (\yb\gamma_\mu\gamma_5
T^z\psi)^2, \label{eq::colorflavor}
\end{eqnarray}
where $(\yb^a\psi^b)^2\equiv \yb^a\psi^b \yb^b \psi^a$, etc., and
$(T^z)_{ij}$ denotes the generators of the gauge group in the
fundamental representation.  The set of fermionic self-interactions
occurring in Eq.\re{equ::truncsym} forms a complete basis. Any other
pointlike four-fermion interaction invariant under
\mbox{$\textrm{SU}(\Nc)$} gauge symmetry and
$\textrm{SU}(\Nf)_{\textrm{L}}\times \textrm{SU}(\Nf)_{\textrm{R}}$
flavor symmetry is reducible by means of Fierz transformations.

It is important to stress that the effective action
\eqref{equ::truncsym} is in the QCD universality class
\cite{Gies:2002hq}, as long as the four-fermion interactions are small
in the UV and thus RG irrelevant. Only nonperturbatively large $(Z_i
\bar\lambda_i)$'s would bring us into another universality class with
\xsb~ of Nambu--Jona-Lasinio (NJL) type, which will not be considered
in this work.

Plugging the truncated effective action \re{equ::truncsym} into the
flow equation, we obtain the following $\beta$ functions for the
dimensionless couplings $\hat{\lambda}_i$ (see \cite{Gies:2003dp}):
\begin{eqnarray}
\!\pat\lm\!\!\!
&=&\!\!\!2\lm\!
    -4v_4 \lFBo\left[ \frac{3}{\Nc}g^2\lm
            -3g^2 \lva \right]
    \label{eq:lm}\\\nonumber
&&\!\!\!\!\!-\frac{1}{8}v_{4}\lFB\left[\frac{12+9\Nc^2}{\Nc^2}g^{4} \right]
\\\nonumber
&&\!\!\!\!\!-8 v_4\lF \Big\{-\Nf\Nc(\lm^2+\lp^2) + \lm^2
\\\nonumber
&&\!\!\quad\quad\quad\quad-2(\Nc+\Nf)\lm\lva
       +\Nf\lp\lsf + 2\lva^2 \Big\},
\end{eqnarray}
\begin{eqnarray}
\!\pat\lp\!\!\!&=&2 \lp\! -4v_4\lFBo \left[
-\frac{3}{\Nc}g^2\lp\right]
    \label{eq:lp}
\\\nonumber
&&\!\!\!\!\!-\frac{1}{8}v_{4}\lFB\left[
    -\frac{12+3\Nc^2}{\Nc^2} g^{4} \right]
\\\nonumber
&&\!\!\!\!\!-8 v_4 \lF \Big\{ - 3\lp^2 - 2\Nc\Nf\lm\lp
\\\nonumber
&&\!\!\quad\quad\quad\quad- 2\lp(\lm+(\Nc+\Nf)\lva)
        + \Nf\lm\lsf
\\\nonumber
&&\!\!\quad\quad\quad\quad+ \lva\lsf
        +\casel{1}{4}\lsf{}^2 \Big\},
    \nonumber\\
\!\pat\lsf\!\!\!
&=&\!\!\!
    2\lsf\! -4v_4 \lFBo \left[6\Cas\, g^2\lsf
    -6g^2\lp \right]\label{eq:lsf}\\\nonumber
&&\!\!\!\!\!-\frac{1}{4} v_4 \lFB \Big[ -\frac{24
    -9\Nc^2}{\Nc}\, g^4  \Big] \\\nonumber
&&\!\!\!\!\! -8 v_4 \lF
     \! \Big\{\! 2\Nc \lsf^2\!  -\! 2\lm\lsf\! - 2\Nf\lsf\lva\!\!
- \!6\lp\lsf\! \Big\},
        \nonumber\\
\!\pat\lva\!\!\!
&=& \!\!\!2 \lva\!-4v_4 \lFBo \left[
\frac{3}{\Nc}g^2\lva -3g^2\lm \right]
    \label{eq:lva} \\
&&\!\!\!\!\!-\frac{1}{8} v_4 \lFB \left[ -\frac{24 - 3\Nc^2}{\Nc}
g^4 \right]
    \nonumber\\
&&\!\!\!\!\! -8 v_4 \lF\!
  \Big\{\! - \!(\Nc+\Nf)\lva^2\! + 4\lm\lva\!
        - \casel{1}{4} \Nf \lsf^2\Big\}.\nonumber
\end{eqnarray}
Here $\Cas=(\Nc^2-1)/(2\Nc)$ is a Casimir operator of the gauge group,
and $v_4=1/(32\pi^2)$. For better readability, we have written all
gauge-coupling-dependent terms in square brackets, whereas fermionic
self-interactions are grouped inside braces.  The threshold functions
$l$ characterize the regulator dependence
\cite{Jungnickel:1995fp}. For the ``optimized'' linear regulator
\cite{Litim:2000ci}, these read
\begin{equation}
\label{optimized}
\lF=\frac{1}{2},\,\,\,\,
\lFBo=1-\frac{\eta_{\textrm{A}}}{6},
\,\,\,\,\lFB=\frac{3}{2}-\frac{\eta_{A}}{6}. 
\end{equation}
Next, let us turn to the running of the gauge coupling. Even though a
non-perturbative estimate can already be obtained within the current
simple truncation \cite{Reuter:1993kw}, a reliable estimate to higher
loops requires the inclusion of many further vertex operators in
the truncation \ref{equ::truncsym}. Instead, we construct the $\beta$
function according to the following line of argument: First, we note
that gauge invariance in the RG scheme can be monitored with the aid
of regulator-dependent Ward-Takahashi identities
\cite{Ellwanger:iz,Reuter:1993kw}. For the present system, these
identities constrain the four-fermion contributions to the running
gauge coupling to be of the form \cite{Gies:2003dp}:
\begin{eqnarray}
\nonumber &&\!\!\!\!\partial_{t}g^2=\breve\eta_{g^2}\, g^2-4 v_4
\lF\, \frac{g^2}{1-2v_4\lF \sum c_i \hat{\lambda}_i} \, \pat \sum c_i
\hat{\lambda}_i, \label{betaeq} \\ &&\!\!\!\!
c_{\sigma}=1+\Nf,\,\,c_{+}=0,\,\,c_{-}=-2,\,\,c_{\textrm{VA}}=-2\Nf,
\end{eqnarray}
where
\begin{eqnarray}
\label{betagauge}
\breve\eta_{g^2}\!\!&=&\!\!\eta_{\textrm{A}}-2\breve\eta_{1}+2\eta_\psi
\\\nonumber
\!\!&=&\!\!-4\vv g^2\left(\beta_{0}+\beta_{1}(2\vv g^2)+\beta_{2}(2\vv
g^2)^2+\ldots\right) 
\end{eqnarray}
is the standard gauge-coupling anomalous dimension, trivially related
to the standard $\beta$ function by $\breve\beta_g^2=\breve\eta_{g^2}
g^2$.$\eta_{\textrm{A}}$, $\eta_\psi$ and $\eta_{1}$ are the gluon,
quark and quark-gluon-vertex anomalous dimensions, respectively,
defined by $\eta_i=-\pat\ln Z_i$; $\eta_{\psi}$, in fact, vanishes in
the present truncation using the Landau gauge. The $\,\,\breve{}\,\,$
symbol denotes those standard contributions which are not generated by
the modified Ward-Takahashi identities; for instance, the difference
between $\eta_1$ and $\breve\eta_1$ is exactly given by the term $\sim
\pat\hat\lambda_i$ in Eq.~\eqref{betaeq}.

Instead of computing $\breve\eta_{g^2}$ in the RG scheme, we use the
four-loop result obtained in the $\bar{\text{MS}}$ scheme
\cite{vanRitbergen:1997va,Czakon:2004bu} as an estimate. For a
qualitative discussion, let us list the first two coefficients (all
four known coefficients are used in the numerics),
\begin{equation}
\label{betacoeff}
\beta_{0}=\frac{11}{3}\Nc-\frac{2}{3}\Nf,\,\,
\beta_{1}=\frac{34\Nc^3+3\Nf-13\Nc^2\Nf}{3\Nc},  
\end{equation}
with $\beta_1$ reversing its sign for
$\Nf>\frac{34\Nc^3}{13\Nc^2-3}$. Since we do not expect that the RG-
and $\bar{\text{MS}}$-scheme $\beta$ functions exactly coincide, this
four-loop approximation introduces an error that is estimated
below.\footnote{Note that the additional terms $\sim \pat
\hat\lambda_i$ in Eq.~\eqref{betaeq} already contribute to the
resulting $\beta$ function at two-loop order. By approximating
$\breve\eta_{g^2}$ by the $\bar{\text{MS}}$ result, we introduce a
potential double counting of diagrams at this and higher order.
However, the $\pat \hat{\lambda}_i$ contributions vanish at the fixed
point.  In addition, we would like to stress that we are using a
mass-dependent regularization scheme, and universality of the second
coefficient in the $\beta$ function does not hold and further
contributions at this order are expected.} It should be stressed at
this point that some crucial properties of the $\beta$ function are not
scheme dependent; in particular, the existence of zeros, i.e., fixed
points, is a universal result which is relevant for the present
problem.

One further technical point needs to be mentioned: whereas the running
gauge coupling is dominated by $\breve\eta_{g^2}$, the threshold
functions $l$ of the fermionic sector require the separate knowledge
of $\eta_{\textrm{A}}$.  As a rough but sufficient estimate, we simply
employ the one-loop relation for the Landau gauge,
$\eta^{\textrm{1-loop}}_{\textrm{A}}
=\frac{13}{22}\breve\eta^{\textrm{1-loop}}_{g^2}$.  Since the
anomalous dimensions turn out to remain numerically small of
${\mathcal{O}}(0.1)$ and the threshold functions only depend on
$\frac{\eta_{\textrm{A}}}{6}$, this additional higher-order
approximation has little effect on the result.\footnote{As a check, we
have used the pure one-loop result
$\eta^{\textrm{1-loop}}_{\textrm{A}}$, which is non-vanishing at the
fixed point. The corresponding results lie well within the estimated
error (shaded region in Fig.~\ref{critnf}).}

\section{Results}\label{results}

The onset of \xsb\ is detected by the four-fermion
part of our model \eqref{equ::truncsym}. Chiral symmetry remains
preserved in the pointlike limit, as long as the fermion
self-interactions remain finite. In turn, \xsb\ is signaled by
a diverging $\hat \lambda_i$. This seeming Landau-pole behavior points
to the formation of chiral condensates, since the $\hat\lambda_i$'s
are inversely proportional to the mass parameter of a Ginzburg-Landau
potential for the order parameter in a (partially) bosonized language,
$\hat\lambda_i\sim 1/m_i^2$. 

\begin{figure}[!t]
\begin{center}
\scalebox{0.75}[0.75]{
\begin{picture}(190,160)(40,0)
\includegraphics[width=9.5cm,height=5cm]{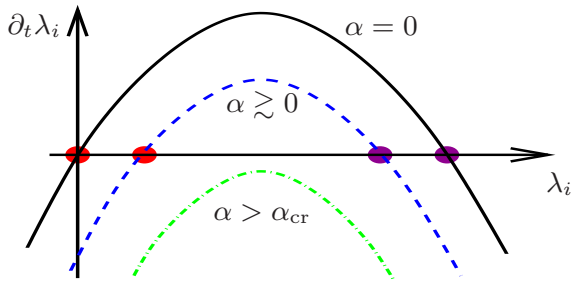}
\Text(-0,+50)[c]{\scalebox{1.6}[1.6]{$\hat\lambda_i$}}
\Text(-265,130)[c]{\scalebox{1.6}[1.6]{$\pat\hat\lambda_i$}} 
\Text(-90,130)[c]{\scalebox{1.6}[1.6]{$\alpha=0$}} 
\Text(-150,90)[c]{\scalebox{1.6}[1.6]{$\alpha\gtrsim 0$}}
\Text(-150,35)[c]{\scalebox{1.6}[1.6]{$\alpha>\alpha_{\text{cr}}$}}
\end{picture}
}
\end{center}
\caption{Sketch of a typical $\beta$ function for the fermionic
  self-interactions $\hat\lambda_i$: at zero gauge coupling,
  $\alpha=0$ (solid curve), the Gau\ss ian fixed point
  $\hat\lambda_i=0$ is IR attractive (the second fixed point at
  $\hat\lambda_i>0$ corresponds to the IR repulsive critical coupling
  of NJL type). For small $\alpha\gtrsim 0$ (dashed curve), the
  fixed-point positions are shifted on the order of $\alpha^2$. For
  gauge couplings larger than the critical coupling
  $\alpha>\alpha_{cr}$ (dot-dashed curve), no fixed points remain and
  the self-interactions quickly grow large, signaling \xsb. We
  emphasize that both fixed-points values remain finite until the
  fixed points eventually vanish at $\alpha_{cr}$.}
\label{fig::parabola}
\end{figure}

Therefore, let us first consider the $\hat\lambda_i$ flow
separately. For vanishing gauge coupling, the flow is solved by
vanishing $\hat\lambda_i$'s, which is the Gau\ss ian fixed point. This
fixed point is IR attractive, implying that these self-interactions
are RG irrelevant for sufficiently small bare couplings, as they
should be. Since the $\hat\lambda_i$ flows are quadratic in the
$\hat\lambda_i$'s corresponding to a parabola, there are 15 further
non-Gau\ss ian fixed points ($2^{(\text{\#}\lambda_i=4)}=16$ in
total).  These are related to the set of critical couplings
$\hat\lambda_{i,\text{cr}}$ beyond which the system is in the
universality class of NJL type, see Fig. \ref{fig::parabola}.

For fixed small gauge coupling, the Gau\ss ian fixed point is shifted
a bit to fixed point values of the order $\hat\lambda_{\ast,i}\sim
g^4$, effectively describing scattering of massless quarks. This fixed
point is still IR attractive, see Fig \ref{fig::parabola}; hence there
is no \xsb\ as long as the gauge coupling remains small enough.

If the gauge coupling becomes larger than a critical value
$\alpha_{\text{cr}}={g^{2}_{\textrm{cr}}}/{4\pi}$, all fixed
points of the $\hat\lambda_i$ flows are removed, such that the
$\hat\lambda_i$'s quickly run into a divergence, signaling \xsb~\cite{Gies:2001nw}.

\begin{figure}[!t]
\begin{center}
\scalebox{0.85}[0.85]{ 
\begin{picture}(190,180)(40,0)
\includegraphics[width=9.5cm]{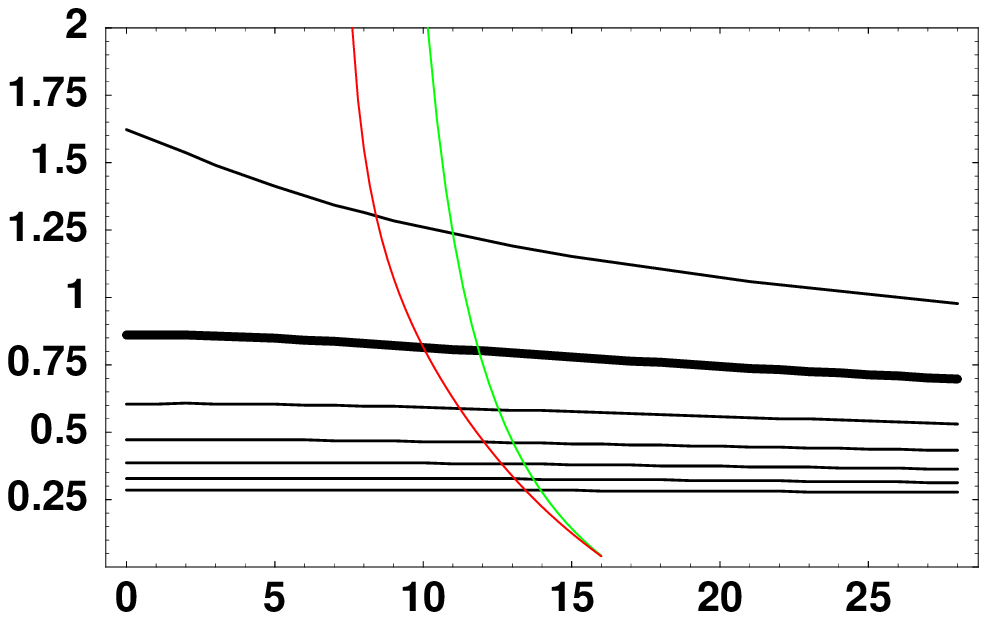}
\Text(-40,-20)[c]{\scalebox{1.6}[1.6]{$\Nf$}}
\Text(-280,150)[c]{\scalebox{1.6}[1.6]{$\alpha_{\textrm{cr}}$}}
\end{picture}
}
\end{center}
\caption{Critical coupling for the four-fermion system. From top to
  bottom, the number of colors increases from $\Nc=2$ to $\Nc=8$
  (thick line corresponds to $\Nc=3$). The red/dark-grey
  (green/light-grey) line shows the fixed-point gauge coupling for
  $\Nc=3$ at four (two) loop.  At the crossing, the critical number of
  flavors can be read off.
}
\label{critcoup}
\end{figure}

In Fig.~\ref{critcoup}, we study the four-fermion system at fixed
gauge coupling. For various numbers of colors $\Nc$, we plot the
critical gauge coupling
$\alpha_{\textrm{cr}}={g^{2}_{\textrm{cr}}}/{4\pi}$ at which the
fermion system becomes critical and exhibits \xsb. For an increasing
number of colors, $\alpha_{\textrm{cr}}$ decreases. The dependence
on the number of flavors is rather weak.

Now, let us switch on the running of the gauge coupling. The
above-mentioned fixed points in the fermionic system become
quasi-fixed points which persist for sufficiently weak gauge coupling;
their values are modulated by the logarithmic increase of the gauge
coupling. If the gauge coupling grows larger than the critical value,
the fermionic quasi-fixed points vanish (the parabola drops below the
$\hat\lambda_i$ axis in Fig.~\ref{fig::parabola}), and the system runs
into the \xsb\ regime. In this way, the IR physics is solely
determined by the gauge coupling, as expected. Whether or not the
system ends up in the \xsb\ regime finally depends on the maximal
value of the gauge coupling. For a large number of flavors,
$\Nf\lesssim\Nf^{\text{a.f.}}$, this value is given by the IR
fixed-point value $\alpha_\ast$, induced by the higher $\beta$
function coefficients. Lowering $\Nf$ increases $\alpha_\ast$,
eventually exceeding $\alpha_{\text{cr}}$ required for \xsb. 

We determine the resulting critical number of flavors
$\Nf^{\text{cr}}$ by comparing the fixed point of the gauge coupling,
$\alpha_\ast\equiv\alpha_\ast(\Nf)$, to $\alpha_{\textrm{cr}}$. The
task of solving the complete coupled set of fixed-point equations
$\partial_{t}g^2=0,\,\partial_{t}\hat{\lambda}=0$ becomes simplified
by the following observation: Eq.~\eqref{betaeq} reveals that the only
contributions to $\pat g^2$ involving $\hat{\lambda}$ are proportional
$\partial_{t}\hat{\lambda}$.  By definition this term vanishes at any
fixed point in the subsystem of four-fermion interactions. Therefore,
the gauge coupling at any fixed point of the complete system is
determined by $\breve\eta_{g^2}=0$, which depends only on $g$.  In
Fig.~\ref{critcoup}, the fixed-point gauge coupling
$\alpha_{\ast}(\Nf)$ is plotted for $\Nc=3$ with a two-
(green/light-grey) and a four-loop (red/dark-grey) result for
$\breve\eta_{g^2}$.

\begin{figure}[!t]
\begin{center}
\scalebox{0.85}[0.85]{ 
\begin{picture}(190,180)(40,0)
\includegraphics[width=9.5cm]{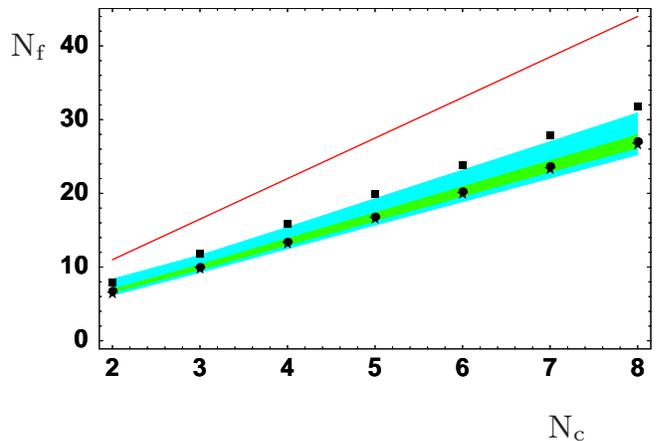}
\Text(-40,-20)[c]{\scalebox{1.6}[1.6]{$\Nc$}}
\Text(-280,150)[c]{\scalebox{1.6}[1.6]{$\Nf$}}
\end{picture}
}
\end{center}
\caption{Critical number of flavors for $\textrm{SU}(\Nc)$ gauge
  theory.  The result based on the four-loop  $\bar{\text{MS}}$ beta
  function is denoted by black circles which lie almost on top of
  stars, representing the three-loop result; black boxes
  correspond to the two-loop beta function.
  The (inner) green/dark-grey shaded region around the four-loop
  result displays our error estimate for the fermionic sector. The
  (outer) turquois/light-grey shaded region shows the approximate
  gluonic error, estimated by a variation of the higher-loop
  coefficients. The red/grey line shows the number of flavors above
  which asymptotic freedom is lost.}
\label{critnf}
\end{figure}

The critical number of flavors $\Nf^{\textrm{cr}}$ can now be read off
from the intersection between this line and the critical coupling
$\alpha_{\textrm{cr}}$ of the four-fermion subsystem for $\Nc=3$
(thick solid line). In the same way, we determine the critical number
of flavors as a function of $\Nc$, as plotted in
Fig.~\ref{critnf}. \footnote{An equivalent way to determine
$\Nf^{\textrm{cr}}(\Nc)$ is to directly solve the flow equations
\eqref{eq:lm}-\eqref{betaeq} with $\hat{\lambda}_{i}(t=0)=0$ and a
small value $g^{2}(t=0)$ as initial conditions. One can then find
$\Nf^{\textrm{cr}}$ by looking for the infimum of all $\Nf$ for which the
four-fermion couplings do not diverge at a finite $t$, (or
alternatively, for the supremum of those which do diverge).}

The regulator dependence of the flow equation offers a possibility for
estimating the error of our quantitative results which is introduced
by the truncation. Exact results for physical quantities like
$\Nf^{\textrm{cr}}$ are universal without any regulator dependence.
However, this universality does not hold for the truncated flow, such
that the regulator dependence of would-be universal quantities
directly translates into an error of the truncation. Let us start with
the fermionic subsystem, 
%
%
for which the point-like approximation of vertices and the use of
``classical'' propagators represents the most severe approximation.
If these truncated nontrivial momentum dependencies of vertices and
propagators were important for the determination of the universal $N_f$,
we would expect a strong regulator dependence within the point-like
approximation. This is because the flow equation is localized in
momentum space and different regulators probe the vertices and
propagators at different momentum shells; consequently, erroneously
truncated momentum dependencies of the vertices generically imply
strong variations of would-be universal quantities for different
regulators.  In this sense, the study of regulator dependencies is an
important if not crucial test of our point-like approximation in the
fermion sector for the determination of $N_f$.
In the present case, the regulator dependence occurs in the
threshold functions $l$. Our pointlike truncation is reminiscent of a
derivative expansion, for which a flow-optimization procedure has led
to the construction of a linear regulator
\cite{Litim:2000ci} used in Eq.~\eqref{optimized}.
Following the stability arguments of \cite{Litim:2000ci},
we consider the results from this regulator as our ``optimized''
predictions. The opposite limit of non-optimized regulators for pointlike
interactions is marked by the sharp cutoff \cite{Litim:2002cf} 
with $\lF=\lFBo=\lFB=1$, neglecting the anomalous
dimension.
Therefore, a conservative error estimate is given by the difference
between the sharp cutoff and the optimized regulator.\footnote{The
sharp cutoff can be considered as the limit of a sequence of powerlike
regulators, the results of which all lie within the estimated error.}
This results in the green/dark-grey shaded region in
Fig.  \ref{critnf}. The resulting error is actually very small; in
particular, we observe strong nontrivial cancellations between the
contributions of different threshold functions to the final
result. Since this exactly matches with the expectation for universal
quantities, it provides evidence for the reliability of the pointlike
truncation for the present purpose.  As a further evidence for the
consistency of the point-like approximation in the symmetric phase, we
note that the gluon as well as the quark anomalous dimension vanish at
the IR fixed point in our approximation, implying that the use of
"classical" propagators is self-consistent here.  Of course, we should
stress that the point-like approximation is expected to break down for
many other questions, in particular, those related to the properties
of the \xsb\ phase.

The largest uncertainty arises from the gauge sector. As a error
estimate, one might use the difference between the various loop orders
in the $\beta$-function contribution $\breve\eta_{g^2}$. As depicted
in Fig. \ref{critnf}, the difference between two and three loops in
the $\bar{\text{MS}}$ scheme is of the order
$\Delta\Nf^{\textrm{cr}}\sim1$. The difference between the three- and
the four-loop result is already much smaller. This suggests that, for
our purposes, the gauge sector is reliably approximated by the
four-loop $\beta$ function. However, as discussed above, the (known)
$\bar{\text{MS}}$ result is not expected to coincide exactly with the
(unknown) result to be obtained with the linear regulator used so far.
For a quantitative estimate of the resulting error, we can perform a
comparison at two-loop level, where the mass-dependent character of
the flow-equation regularization already introduces scheme
dependencies. On a quantitative level, this introduces differences on
the 10\% level \cite{Reuter:1997gx}.\footnote{For an exact computation
  of the universal two-loop coefficient within the functional RG
  approach, see \cite{Pawlowski:2001df}.} For the 3- and 4-loop
  coefficients we assume larger uncertainties on the 30\% and 50\%
  level, respectively.  This results in an error depicted by the
turquois/light-grey shaded region in Fig. \ref{critnf}.  Quadratically
adding both uncertainties, the resulting error is dominated by the
gluonic sector. 

It has been argued \cite{tHooft} that a regularization scheme exists
in which the two-loop $\beta$ function for the gauge coupling is
exact. We stress that this does not imply that the two-loop curve in
Fig.~\ref{critnf} is exact, since this particular unknown
regularization scheme may involve a regulator with strong quantitative
influence on the fermionic sector (possibly requiring operators beyond
the pointlike limit). Hence, it is well possible that the resulting
curve of that special scheme is again close to the estimated four-loop
result in our scheme.

It is worth pointing out that the total estimated error is
significantly smaller than the distance to the line
$\Nf^{\textrm{a.f.}}(\Nc)$ where asymptotic freedom is lost. This
strongly confirms the existence of a phase of many-flavor QCD which is
asymptotically free, but has no \xsb\ and no strict confinement in the
IR.  

\section{Summary and Conclusions}\label{summary}
We have used flow equations to determine the critical number of
flavors $\Nf^{\textrm{cr}}$ separating the phases with and without
\xsb. Our truncation includes all pointlike four-fermion interactions
allowed by gauge and chiral symmetry which are generated by ladder as
well as non-ladder processes; thereby, our result represents a
significant improvement over conventional approximation
techniques.  Gauge symmetry is monitored by modified Ward-Takahashi
identities, resulting in an additional four-fermion contribution to the
flow of the gauge coupling. We have demonstrated that these terms have
no effect on $\Nf^{\textrm{cr}}$, owing to the structure of the
modified Ward-Takahashi identity (by contrast, these terms are of
quantitative importance, for instance, for the Landau-pole problem of
QED \cite{Gies:2004hy}).

Our findings confirm the existence of a sizable parameter region
$\Nf^{\textrm{cr}}<\Nf<\Nf^{\textrm{a.f.}}$ where chiral symmetry
remains unbroken, in agreement with earlier results
\cite{Miransky:1996pd,Appelquist:1996dq,Harada:2003dc,Iwasaki:2003de,Sannino:1999qe}.
Our result for the critical number of flavors, for instance, in SU(3)
gauge theory, is $\Nf^{\text{cr}}=10.0\pm
  0.29\text{(fermion)}\genfrac{}{}{0pt}{}{+1.55}{-0.63}\text{(gluon)}$. The
  errors result from the fermionic and gluonic sectors of our
  truncation, respectively, the quantitative influence of which we
  have estimated from the regulator dependence of universal
  quantities.

Our restriction to a truncation in terms of quark and gluon fields
with pointlike interaction does not facilitate a study of the \xsb\
regime; for this, the inclusion of bosonic composite fields along the
lines of \cite{Gies:2001nw,Gies:2002hq,Jaeckel:2002rm,Jaeckel:2003uz}
represents a powerful tool.

Nevertheless, important aspects of the unusual nature of the phase
transition can already be read off from the fermionic sector: as long
as the system is in the symmetric phase,
$\Nf^{\text{cr}}<\Nf<\Nf^{\text{a.f.}}$, the fermionic
self-interactions approach fixed-point values in the IR which are
finite numbers. Even arbitrarily close to the phase transition, these
numbers remain finite and are elements of a compact region in parameter
space, $\hat\lambda_i|_{k\to 0} \in \Omega_{\hat\lambda}$. By
contrast, the self-interactions diverge in the broken phase where all
fixed points have vanished. As a consequence, an infinite region in
parameter space outside $\Omega_{\hat\lambda}$ remains
inaccessible. Varying $\Nf$ continuously across the phase transition
coming from the symmetric side, the fermionic self-interactions jump
discontinuously at $\Nf=\Nf^{\text{cr}}$. Now, the $\hat\lambda_i$'s
are inversely proportional to the mass parameter of a Ginzburg-Landau
potential for the order parameter in a (partially) bosonized language,
$\hat\lambda_i\sim 1/m_i^2$; hence, our observation in the fermionic
sector agrees with the observations of
\cite{Miransky:1996pd,Appelquist:1996dq,Chivukula:1996kg} that the
phase transition is not conventionally second order, even though the
chiral order parameter is known to change continuously across the
chiral phase transition \cite{Gies:2002hq}. Our work thus reveals
further aspects of the nature of this type of zero-temperature phase
transitions which may find further application in related systems such
as QED${}_3$ or models of electroweak symmetry breaking.

\acknowledgments{J.J.~would like to thank the ITP in Heidelberg for
its hospitality and A. Ringwald for useful comments. 
H.G.~is grateful to J.M. Pawlowski for discussions and acknowledges support by the Deutsche
Forschungsgemeinschaft (DFG) under contract Gi 328/1-3 (Emmy-Noether
program).}


\begin{thebibliography}{99}
\bibitem{Banks:1981nn}
T.~Banks and A.~Zaks,
Nucl.\ Phys.\ B {\bf 196}, 189 (1982).

\bibitem{Miransky:1996pd}
V.~A.~Miransky and K.~Yamawaki,
Phys.\ Rev.\ D {\bf 55}, 5051 (1997)
[Erratum-ibid.\ D {\bf 56}, 3768 (1997)]
[hep-th/9611142].

\bibitem{Appelquist:1996dq}
T.~Appelquist, J.~Terning and L.~C.~R.~Wijewardhana,
Phys.\ Rev.\ Lett.\  {\bf 77}, 1214 (1996)
[hep-ph/9602385];\\
T.~Appelquist, A.~Ratnaweera, J.~Terning and L.~C.~R.~Wijewardhana,
Phys.\ Rev.\ D {\bf 58}, 105017 (1998)
[hep-ph/9806472].

\bibitem{Harada:2003dc}
M.~Harada, M.~Kurachi and K.~Yamawaki,
Phys.\ Rev.\ D {\bf 68}, 076001 (2003)
[hep-ph/0305018].

\bibitem{Iwasaki:2003de}
Y.~Iwasaki, K.~Kanaya, S.~Kaya, S.~Sakai and T.~Yoshie,
Phys.\ Rev.\ D {\bf 69}, 014507 (2004)
[hep-lat/0309159].

\bibitem{Sannino:1999qe}
F.~Sannino and J.~Schechter,
Phys.\ Rev.\ D {\bf 60}, 056004 (1999)
[hep-ph/9903359].

\bibitem{Appelquist:1997dc}
  T.~Appelquist and S.~B.~Selipsky,
  Phys.\ Lett.\ B {\bf 400}, 364 (1997)
  [hep-ph/9702404];\\
M.~Velkovsky and E.~V.~Shuryak,
Phys.\ Lett.\ B {\bf 437}, 398 (1998)
[hep-ph/9703345].

\bibitem{Harada:2000kb}
  M.~Harada and K.~Yamawaki,
  Phys.\ Rev.\ Lett.\  {\bf 86} (2001) 757
  [hep-ph/0010207].

\bibitem{Chivukula:1996kg}
R.~S.~Chivukula,
Phys.\ Rev.\ D {\bf 55}, 5238 (1997)
[hep-ph/9612267].

\bibitem{Wetterich:1993yh}
C.~Wetterich,
Phys.\ Lett.\ B {\bf 301}, 90 (1993);
Nucl.\ Phys.\ B {\bf 352}, 529 (1991);
Z.\ Phys.\ C {\bf 48}, 693 (1990).
%

\bibitem{Gies:2003dp}
H.~Gies, J.~Jaeckel and C.~Wetterich,
Phys.\ Rev.\ D {\bf 69} (2004) 105008
[hep-ph/0312034].

\bibitem{Ellwanger:1995qf}
U.~Ellwanger, M.~Hirsch and A.~Weber,
Z.\ Phys.\ C {\bf 69}, 687 (1996)
[hep-th/9506019];\\
D.~F.~Litim and J.~M.~Pawlowski,
Phys.\ Lett.\ B {\bf 435}, 181 (1998)
[hep-th/9802064].

\bibitem{Gies:2002hq}
H.~Gies and C.~Wetterich,
Phys.\ Rev.\ D {\bf 69}, 025001 (2004)
[hep-th/0209183].

\bibitem{Jungnickel:1995fp}
D.~U.~Jungnickel and C.~Wetterich,
Phys.\ Rev.\ D {\bf 53}, 5142 (1996)
[hep-ph/9505267].

\bibitem{Litim:2000ci}
  D.~F.~Litim,
  Phys.\ Lett.\ B {\bf 486} (2000) 92
  [hep-th/0005245]; 
  Phys.\ Rev.\ D {\bf 64} (2001) 105007
  [hep-th/0103195].

\bibitem{Reuter:1993kw}
M.~Reuter and C.~Wetterich,
Nucl.\ Phys.\ B {\bf 417}, 181 (1994).
%

\bibitem{Ellwanger:iz}
U.~Ellwanger,
Phys.\ Lett.\ B {\bf 335}, 364 (1994)
[hep-th/9402077];\\
%
F.~Freire, D.~F.~Litim and J.~M.~Pawlowski,
Phys.\ Lett.\ B {\bf 495}, 256 (2000)
[hep-th/0009110].
%

\bibitem{vanRitbergen:1997va}
T.~van Ritbergen, J.~A.~M.~Vermaseren and S.~A.~Larin,
Phys.\ Lett.\ B {\bf 400} (1997) 379
[hep-ph/9701390].

\bibitem{Czakon:2004bu}
  M.~Czakon,
  Nucl.\ Phys.\ B {\bf 710} (2005) 485
  [hep-ph/0411261].

\bibitem{Gies:2001nw}
  H.~Gies and C.~Wetterich,
  Phys.\ Rev.\ D {\bf 65} (2002) 065001
  [hep-th/0107221].

\bibitem{Litim:2002cf}
  D.~F.~Litim,
  Nucl.\ Phys.\ B {\bf 631} (2002) 128
  [hep-th/0203006].

\bibitem{Reuter:1997gx}
M.~Reuter and C.~Wetterich,
Phys.\ Rev.\ D {\bf 56}, 7893 (1997)
[hep-th/9708051];\\
H.~Gies,
Phys.\ Rev.\ D {\bf 66}, 025006 (2002)
[hep-th/0202207].

\bibitem{Pawlowski:2001df}
J.~M.~Pawlowski,
Int.\ J.\ Mod.\ Phys.\ A {\bf 16}, 2105 (2001).

\bibitem{tHooft}
G.~'t Hooft, in {\em Recent Developments in Gauge Theories}, NATO
Advanced Study Inst., Ser. B, v.59, eds. G.~t Hooft et al., Plenum
Press, New York (1980).

\bibitem{Gies:2004hy}
H.~Gies and J.~Jaeckel,
Phys.\ Rev.\ Lett.\  {\bf 93}, 110405 (2004)
[hep-ph/0405183].


\bibitem{Jaeckel:2002rm}
  J.~Jaeckel and C.~Wetterich,
  Phys.\ Rev.\ D {\bf 68} (2003) 025020
  [hep-ph/0207094].

\bibitem{Jaeckel:2003uz}
  J.~Jaeckel,
  hep-ph/0309090.
\end{thebibliography}
\end{document}